\documentclass[twocolumn,showpacs,preprintnumbers,amsmath,amssymb]{revtex4}

\usepackage{graphicx}
\usepackage{dcolumn}
\usepackage{./simplewick}
\usepackage{bm}
\usepackage{hyperref}
\usepackage{color}
\begin{document}

\title{Properties from relativistic coupled-cluster without 
       truncation: hyperfine constants of $^{25}{\rm Mg}^+$,
       $^{43}{\rm Ca}^+ $, $^{87}{\rm Sr}^+ $ and $^{137}{\rm Ba}^+$}
\author{B. K. Mani and D. Angom}
% \email{bkmani@prl.res.in}
% \email{angom@prl.res.in}
\affiliation{Physical Research Laboratory,
             Navarangpura-380009, Gujarat, 
             India}%\textbackslash\textbackslash}

\begin{abstract}
   We demonstrate an iterative scheme for coupled-cluster properties 
calculations without truncating the dressed properties operator. For
validation, magnetic dipole hyperfine constants of alkaline Earth ions are 
calculated with relativistic coupled-cluster and role of electron correlation 
examined. Then, a detailed analysis of the higher order terms is carried out. 
Based on the results, we arrive at an optimal form of the dressed operator. 
Which we recommend for properties calculations with relativistic 
coupled-cluster theory.
\end{abstract}


%% 31.15.bw: Coupled-cluster theory
%% 32.10.Fn: Fine and hyperfine structure
%% 31.15.vj: Electron correlation calculations for atoms and ions: 
%%           excited states
%% 31.15.am: Relativistic configuration interaction (CI) and many-body 
%%           perturbation calculations

\maketitle

\maketitle

%%%%%%%%%%%%%%%%%%%%%%%%%%%%%%%%%%%%%%%%%%%%%%%%%%%%%%%%%%%%%%%%%%%%%%%%%%%%%%
%%%%%%%%%%%              Introduction                    %%%%%%%%%%%%%%%%%%%%%
%%%%%%%%%%%%%%%%%%%%%%%%%%%%%%%%%%%%%%%%%%%%%%%%%%%%%%%%%%%%%%%%%%%%%%%%%%%%%%
\section{Introduction}

  Coupled-cluster theory, first developed in nuclear many body physics
\cite{coester-58,coester-60}, is considered one of the best many body theory. 
In recent times, it has been used with great success in nuclear 
\cite{hagen-08}, atomic \cite{nataraj-08,pal-07}, molecular \cite{isaev-04} 
and condensed matter \cite{bishop-09} calculations. In atoms it is equivalent 
to incorporating electron correlation effects to all order. It has been used 
extensively in precision atomic properties and structure calculations. These 
include atomic electric dipole moments \cite{nataraj-08,latha-09}, parity 
nonconservation \cite{wansbeek-08}, hyperfine structure constants 
\cite{pal-07,sahoo-09} and electromagnetic transition properties 
\cite{thierfelder-09,sahoo-09a}.

  Despite the remarkable developments and numerous calculations based
on relativistic coupled-cluster theory. Hitherto, a systematic analysis
of the properties calculations with coupled-cluster wave functions is lacking.
This issue arises from the fact that, the expression for properties with
coupled-cluster wave functions is a non terminating series.  In this paper
we demonstrate an iterative scheme to calculate properties without truncation. 
Such a study is essential and timely as precision atomic calculations, in 
several instances, complement precision atomic experiments. These have 
direct bearing on, to mention a few, fundamental physics and new technology.

 To test and validate the scheme we employ open shell coupled-cluster theory
\cite{Mukherjee-79,Salomonson-80,Lindgren-85} and calculate the magnetic 
dipole hyperfine constants of alkaline Earth ions $^{25}{\rm Mg}^+$, 
$^{43}{\rm Ca}^+$, $^{87}{\rm Sr}^+$ and $^{137}{\rm Ba}^+$. We have selected 
these ions as these are potential 
candidates for ongoing or proposed novel experiments. In addition, there is 
a large variation in the role of electron correlation among the ions and 
states. The ground state hyperfine constant of Mg$^+$ is well studied with 
ion trapping techniques \cite{Itano-81}. The clock states of, the next ion 
in the group, $^{43}{\rm Ca}^+$ were recently employed for high-fidelity 
entanglement \cite{Kirchmair-09}. A crucial step in quantum information 
processing. Then, single trapped $^{87}{\rm Sr}^+ $ is a suitable frequency 
standard \cite{Barwood-03}. These are application oriented precision 
experiments. The other fascinating prospect is observation of parity 
nonconservation in a single $^{137}{\rm Ba}^+$ \cite{Fortson-93}. In all of 
these endeavours, hyperfine interaction is involved.  For this reason, 
several theoretical calculations have examined the role of electron 
correlations to the hyperfine constants of these ions. These provide a 
wealth of data for comparative study. In addition to magnetic hyperfine
constant, we also compute the excitation energies  of the low lying states.
This is to verify the quality of the single particle wave function we use.

  The paper is divided into seven sections. In the next section, that is 
Section.II,  we give a brief description of single valence coupled-cluster 
theory. Then, Section.III is a short writeup on hyperfine interaction and
how it is calculated with relativistic coupled-cluster. Section.IV forms
the core of the paper, where we explain our iterative scheme to calculate 
properties with relativistic coupled-cluster to all order. The details of the
numerical methods and schemes used in the present work are provided in 
Section.V. And then we present our results in Section.VI. Finally, in 
Section.VII we make concluding remarks, which may serve as guideline for any 
properties calculations with relativistic coupled-cluster theory.
In the paper, all the calculations and mathematical expressions are in atomic
units ($e=\hbar=m_e=1$).

%%%%%%%%%%%%%%%%%%%%%%%%%%%%%%%%%%%%%%%%%%%%%%%%%%%%%%%%%%%%%%%%%%%%%%%%%%%%%%
%%%%%               The coupled-cluster method                          %%%%%%
%%%%%%%%%%%%%%%%%%%%%%%%%%%%%%%%%%%%%%%%%%%%%%%%%%%%%%%%%%%%%%%%%%%%%%%%%%%%%%

\section{Single valence coupled-cluster theory} 

\begin{figure}[h]
  \includegraphics[width = 4.0cm]{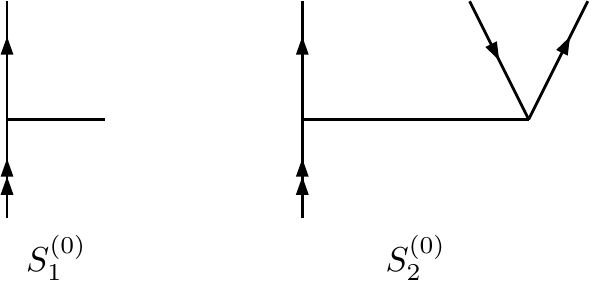}
  \caption{Diagrammatic representation of open shell cluster operators. The
           orbital lines with double arrows indicate valence and single
           up (down) arrow indicate particle (hole) states.}
  \label{fig-s1s2}
\end{figure}

  For completeness and easy reference of the working equations, we provide a
condensed overview of the single valence coupled-cluster theory. Readers are 
referred to Ref. \cite{Lindgren-85} for a detailed exposition of the theory.
In the Fock space coupled-cluster theory of single valence systems, the 
correlated wave function is calculated in two steps. First, the cluster 
operators of the core electrons or the closed-shell part $T$ is evaluated 
from the reference state $|\Phi_0\rangle$. Second, the cluster operators of 
the valence shells $S$ is evaluated and the reference state is
\begin{equation}
  |\Phi_v\rangle = a_v^{\dagger}|\Phi_0\rangle.
\end{equation}
The coupled-cluster wave function of the open shell system is
\begin{equation}
  |\Psi_v\rangle = e^{T + S}|\Phi_v\rangle .
\end{equation}
For single valence system $ e^S=1 + S$, the higher order terms in the 
exponential do not contribute. Then
\begin{equation}
  |\Psi_v\rangle = e^T(1 + S )|\Phi_v\rangle .
  \label{ccwaveopen}
\end{equation}
For an $N$ electron atom, the cluster operators are
\begin{equation}
  T= \sum_{i=1}^{N-1} T_i, \;\;{\rm and} \;\;
  S = \sum_{i = 1}^N S_i.
  \label{t}
\end{equation}
Here the summation index of the $T$ is up to the $N-1$ core electrons, where 
as $S$ is up to $N$ to include the valence electron. However, single and double 
are the most dominant, in coupled-cluster singles and doubles (CCSD) 
approximation $T= T_1  + T_2$ and $S=S_1 + S_2$. In the second quantized 
representation, for the closed-shell part
\begin{equation}
  T_1 = \sum_{a, p}t_a^p a_p^{\dagger}a_a, \;\; {\rm and}\;\;
  T_2 = \frac{1}{2!}\sum_{a, b, p, q}t_{ab}^{pq}
  a_p^{\dagger}a_q^{\dagger}a_ba_a.
\end{equation}
Similarly, for the valence shell 
\begin{equation}
  S_1 = \sum_{p}s_v^p a_p^{\dagger}a_v, \;\; {\rm and}\;\;
  S_2 = \sum_{a, p, q}s_{va}^{pq}
  a_p^{\dagger}a_q^{\dagger}a_aa_v.
\end{equation}
Here, $t_{\cdots}^{\cdots}$ and $s_{\cdots}^{\cdots}$ are the cluster 
amplitudes. The indexes $abc\ldots$ ($pqr\ldots$) represent core (virtual)
states and $vwx\ldots$ represent valence states. The operators $T_1$ ($S_1$ )
and $T_2$ ($S_2$) give single and double replacements after operating on the 
closed-(open-)shell reference states. The diagrammatic representation of 
$S$ are shown in Fig.\ref{fig-s1s2}.

The atomic state $|\Psi_v\rangle$ satisfy the eigen value equation
\begin{equation}
  H|\Psi_v\rangle = E_v |\Psi_v\rangle ,
\end{equation}
where $H$ is the atomic Hamiltonian and $E_v$ is the exact eigen energy of the 
atomic state. Applying $e^{-T} $ on the above equation, we get
\begin{equation}
  \bar H (1 + S)|\Phi_v\rangle = E_v(1 + S)|\Phi_v\rangle,
  \label{Hbar}
\end{equation}
where
\begin{equation}
  \bar H=H+\{\contraction[0.5ex]{}{H}{}{T}HT\} +
  \frac{1}{2!}\{\contraction[0.5ex]{}{H}{}{T}
  \contraction[0.8ex]{}{H}{T}{T} HTT\} +
  \frac{1}{3!}\{\contraction[0.5ex]{}{H}{}{T}
  \contraction[0.8ex]{}{H}{T}{T}
  \contraction[1.1ex]{}{H}{TT}{T}HTTT\}  
  + \frac{1}{4!}\{\contraction[0.5ex]{}{H}{}{T}\contraction[0.8ex]{}{H}{T}{T}
  \contraction[1.1ex]{}{H}{TT}{T}\contraction[1.4ex]{}{H}{TTT}{T} HTTTT\},
  \label{Hnbarcont}
\end{equation}
is the dressed Hamiltonian and $\{\cdots \}$ denotes normal ordering of the 
operators and $ \{\contraction[0.5ex]{}{A}{\cdots}{B}A\cdots B\}$ represents
contraction between two operators $A$ and $B$. The cluster amplitude equations 
of the singles and doubles are obtained after projecting Eq.(\ref{Hbar}) on 
singly and doubly replaced states $\langle\Phi_v^p|$ and 
$\langle\Phi_{va}^{pq}|$. From Wick's theorem and the normal ordered form of 
Hamiltonian ($H_N = H - \langle\Phi_v|H|\Phi_v\rangle = H - E_v^{(0)}$),
we get after the projection
\begin{eqnarray}
  \langle \Phi_v^p|\bar H_N \! +\! \{\contraction[0.5ex]{\bar}{H}{_N}{S}
  \bar H_N S_1\} \! + \! \{\contraction[0.5ex]{\bar}{H}{_N}{S}
  \bar H_N S_2\} |\Phi_v\rangle \!\!\! &= & \!\!\!
  \Delta E_v^{\rm att}\langle\Phi_v^p|S_1|\Phi_v\rangle ,
  \label{ccsingles}     \\
  \langle \Phi_{va}^{pq}|\bar H_N +\{\contraction[0.5ex]{\bar}{H}{_N}{S}
  \bar H_N S_1\}+\{\contraction[0.5ex]{\bar}{H}{_N}{S}
  \bar H_N S_2\} |\Phi_v\rangle &=&  
  \Delta E_v^{\rm att}\langle\Phi_{va}^{pq}|S_2|\Phi_v\rangle.
  \label{ccdoubles}
\end{eqnarray}
In these equations, $\Delta E_v^{\rm att}$ is the valence correlation energy.
It is defined as
\begin{equation}
   \Delta E_v^{\rm att} = \Delta E_v^{N, {\rm corr}} - 
               \Delta E_v^{N-1, {\rm corr}},
\end{equation}
where $\Delta E_v^{N, {\rm corr}}$ and $\Delta E_v^{N-1, {\rm corr}}$ are 
the total and core correlation energies respectively.
%%%%%%%%%%%%%%%%%%%%%%%%%%%
% cluster operator diagrams 
%%%%%%%%%%%%%%%%%%%%%%%%%%%
The right members in Eq.(\ref{ccsingles}-\ref{ccdoubles}) are what 
distinguishes the open shell coupled-cluster theory from that of the 
closed-shell. These are the equivalent of the folded diagrams in the 
many-body perturbation theory (MBPT) of open shell systems. 

%%%%%%%%%%%%%%%%%%%%%%%%%%%%%%%%%%%%%%%%%%%%%%%%%%%%%%%%%%%%%%%%%%%%%%%%%%%%%%%
%%%%%               Subsection: Energy eigenvalue                       %%%%%%%
%%%%%%%%%%%%%%%%%%%%%%%%%%%%%%%%%%%%%%%%%%%%%%%%%%%%%%%%%%%%%%%%%%%%%%%%%%%%%%%

\subsection{Energy eigenvalue}
To obtain the energy eigenvalue $E_v$ of the state $|\Psi_v\rangle$, 
project Eq. (\ref{Hbar}) on the state $\langle \Phi_v|$. Then
\begin{eqnarray}
  \langle\Phi_v|\bar H (1 + S)|\Phi_v\rangle= E_v,
  \label{exactE}
\end{eqnarray}
here we have used $\langle\Phi_v|S|\Phi_v\rangle = 0$. Using the normal 
ordered Hamiltonian, defined earlier, Eq. (\ref{exactE}) can be written as
\begin{equation}
  \langle\Phi_v|\left [ \bar H_N + E_v^{(0)}\right ] 
        (1 + S)|\Phi_v\rangle = E_v.
\end{equation}
From Wick's theorem 
\begin{equation}
  \langle\Phi_v|\left [ \bar H_N + \{\contraction[0.5ex]{\bar}{H}{_N}{S}
      \bar H_N S\} \right ] |\Phi_v\rangle = \Delta E_v^{N, {\rm corr}}. 
\end{equation}
The attachment energy is the difference in the exact energy of the $N$- and 
$(N-1)$-electron state (closed-shell). In terms of correlation energies,
attachment energy
\begin{equation}
  E_v^{\rm att} = \Delta E_v^{N, {\rm corr}} - 
                  \Delta E_v^{N-1, {\rm corr}} + \epsilon_v,
\end{equation}
where $\epsilon_v$ is the single electron energy of the valence electron.
%%%%%%%%%%%%%%%%%%%%%%%%%%%%
% attachment energy diagrams
%%%%%%%%%%%%%%%%%%%%%%%%%%%%
\begin{figure}[h]
  \includegraphics[width = 8cm]{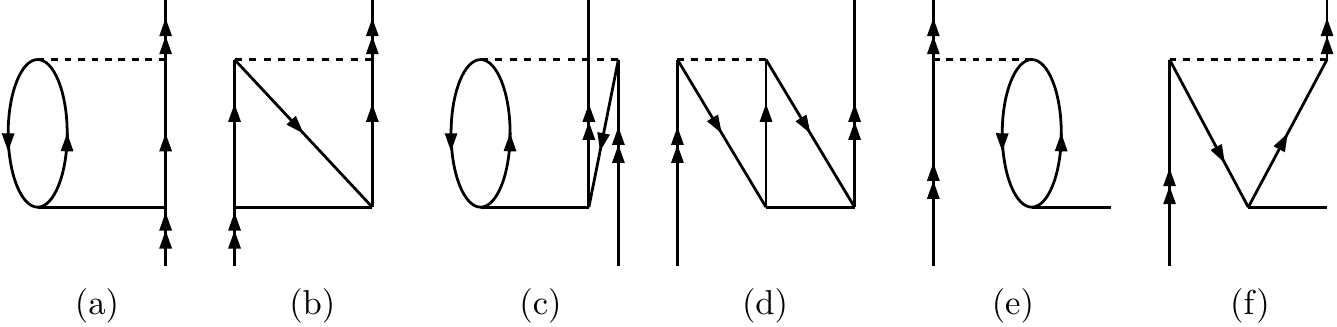}
  \caption{Diagrams which contribute to $\Delta E_v^{\rm att}$. 
           The dashed line represent the residual Coulomb interaction.}
  \label{fig-att-diag}
\end{figure}
From the closed-shell coupled-cluster theory, the correlation energy
$\Delta E_v^{N-1, {\rm corr}}$ have contribution from the closed diagrams.
The right members in the amplitude equations 
Eq.(\ref{ccsingles}-\ref{ccdoubles}) absorb this correlation energy as
$ \langle \Phi_{\cdots}^{\cdots}|\{ \bar H_N S\}|\Phi_v\rangle$ is 
equivalent to 
$\Delta E_v^{N-1, {\rm corr}}\langle\Phi_{\cdots}^{\cdots}|S|\Phi_v\rangle$.
Then the diagrams which contribute to $\Delta E_v^{\rm att}$ are the ones 
shown in Fig.\ref{fig-att-diag}.

%%%%%%%%%%%%%%%%%%%%%%%%%%%%%%%%%%%%%%%%%%%%%%%%%%%%%%%%%%%%%%%%%%%%%%%%%%%%%%%
%%%%%               Subsection: Multiple valence shells                 %%%%%%%
%%%%%%%%%%%%%%%%%%%%%%%%%%%%%%%%%%%%%%%%%%%%%%%%%%%%%%%%%%%%%%%%%%%%%%%%%%%%%%%

\subsection{Multiple valence shells}

  It is relatively straight forward to calculate, from the single valence 
CCSD theory described, the ground state wave function and energy. 
Then the entire single particle basis space consist of one valence orbital,
and the remaining are core (occupied) and virtual (unoccupied). However, to 
calculate excitation energies, the excited atomic states and eigenvalues 
must be calculated. The trivial way is to solve the CCSD equations of each 
atomic states, ground and excited, separately. For example, to evaluate the 
$5d\; ^2D_{3/2}$ excitation energy of Ba$^+$ ion, the ground state 
$|6s\; ^2S_{1/2}\rangle$ and the excited state $|5d\;^2D_{3/2}\rangle$ must be 
calculated. Which translates to solving two sets of CCSD equations with 
$a^\dagger_{6s}|{\rm Ba}^{2+}\rangle$
and $a^\dagger_{5d_{3/2}}|{\rm Ba}^{2+}\rangle$ as reference states. Here, 
$|{\rm Ba }^{2+} \rangle$  is the closed shell Ba$^{2+}$ reference state. 

 A better approach is to solve the ground and excited states CCSD equations
in a single calculation. Then the theory is multi reference in nature and
the cluster equations of different states are coupled. In the present case,
we choose the model space to consist of one state of specific $J$ and parity.
Hence we do not have to invoke a full fledged multi reference 
coupled-cluster theory.

%%%%%%%%%%%%%%%%%%%%%%%%%%%%%%%%%%%%%%%%%%%%%%%%%%%%%%%%%%%%%%%%%%%%%%%%%%%%%%%
%%%%%                    Section: Properties calculation                 %%%%%%
%%%%%%%%%%%%%%%%%%%%%%%%%%%%%%%%%%%%%%%%%%%%%%%%%%%%%%%%%%%%%%%%%%%%%%%%%%%%%%%

\section{Properties calculation}

%%%%%%%%%%%%%%%%%%%%%%%%%%%%%%%%%%%%%%%%%%%%%%%%%%%%%%%%%%%%%%%%%%%%%%%%%%%%%%%
%%%%%             Subsection: Hyperfine structure constants               %%%%%
%%%%%%%%%%%%%%%%%%%%%%%%%%%%%%%%%%%%%%%%%%%%%%%%%%%%%%%%%%%%%%%%%%%%%%%%%%%%%%%

\subsection{Hyperfine Structure Constants}

  The hyperfine interaction $H_{\rm hfs}$ is the coupling of the nuclear 
electromagnetic moments to the electromagnetic field of the electrons. This 
causes splitting of the atomic levels and total angular momentum $F$ is the 
conserved quantity. The atomic states are then $|(IJ)FM_F\rangle$, here $I$ 
and $J$ are the nuclear spin and total electronic angular momentum 
respectively. The general form of the interaction is \cite{Charles-55}
\begin{equation}
  H_{\rm hfs} = \sum_i\sum_{k, q}(-1)^q t^k_q(\hat {\bf r}_i) T^k_{-q},
  \label{hfs_ham}
\end{equation}
where $t^k_q(\bm{r})$ and $T^k_{q}$ are irreducible tensor operators of rank
$k$ effective in the electron and nuclear spaces respectively. From 
the parity selection, only even and odd values of $k$ are allowed for 
electric and magnetic interactions respectively. For the magnetic dipole 
interaction ($k = 1$), the explicit form of the tensor operators are
\begin{equation}
  t^1_q({\bf r}) = \frac{-i\sqrt{2}[{\bm \alpha}\cdot{\bf C}_1
                        (\hat {\bf r})]_q} {cr^2},
                        \;\;\;\;{\rm and}\;\;\;\;\;
  T^1_q = \mu_q.
  \label{hfs_magnetic}
\end{equation}
Here,  ${\bf C}_1(\hat {\bf r})$ is a rank one tensor operator in electron 
space and $\mu _q$ is a component of $\bm \mu$, the nuclear magnetic moment
operator. Then the nuclear moment is the expectation value of $\bm \mu$ in the 
stretched state $ \mu = \langle II|\mu_0|II\rangle$. Parameters which 
represents the hyperfine splitting are the hyperfine structure constants. For
one valence atoms, the magnetic dipole hyperfine structure constant 
\begin{equation}
  a = \frac{g_I\mu_N}{\sqrt{j_v(j_v+1)(2j_v+1)}}
      \langle n_v\kappa_v ||t^1||n_v\kappa_v \rangle.
      \label{hfs_mdipole}
\end{equation}
Here, $g_I$ $(\mu = g_II\mu_N)$ is the gyromagnetic ratio and 
$\mu_N$ is the nuclear magneton.

%%%%%%%%%%%%%%%%%%%%%%%%%%%%%%%%%%%%%%%%%%%%%%%%%%%%%%%%%%%%%%%%%%%%%%%%%%%%%%
%%%%       Subsection:Hyperfine constants from coupled-cluster          %%%%%%
%%%%%%%%%%%%%%%%%%%%%%%%%%%%%%%%%%%%%%%%%%%%%%%%%%%%%%%%%%%%%%%%%%%%%%%%%%%%%%

\subsection{Hyperfine constants from coupled-cluster} 

\begin{center}
\begin{figure}
  \includegraphics[width = 8.0cm]{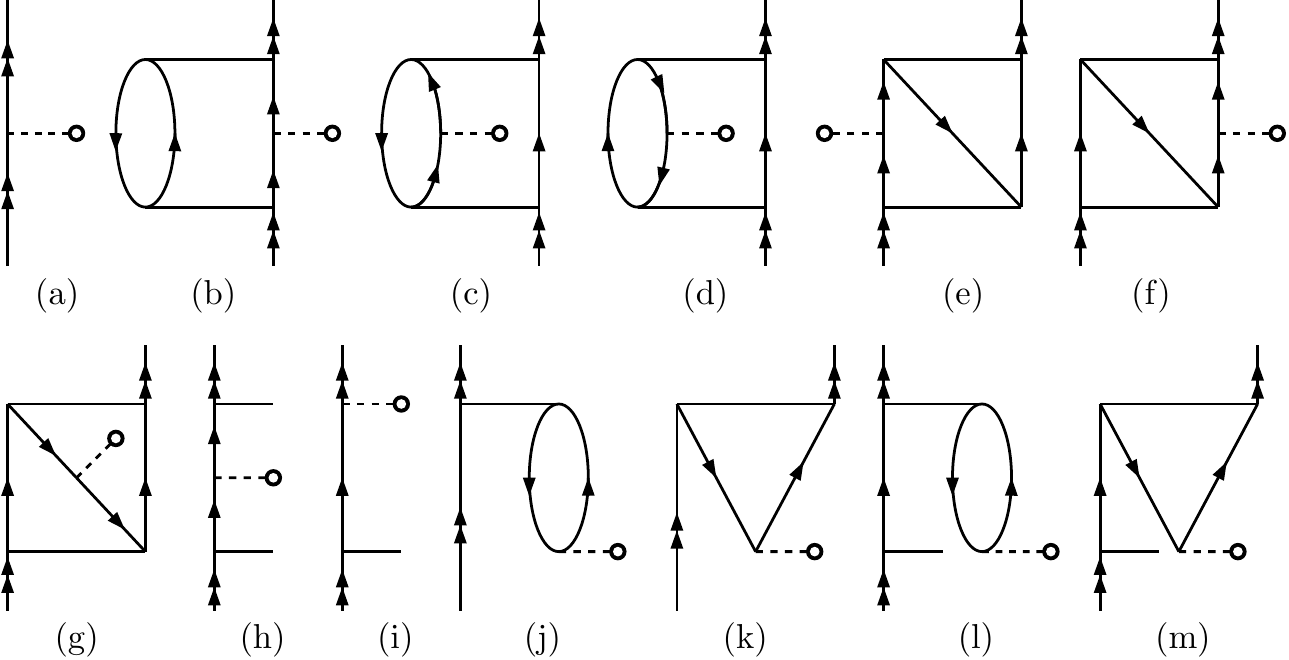}
  \caption{Selected leading diagrams contributing to the hyperfine 
           structure constants in Eq.(\ref{hfs_num}).
           The dashed lines terminated with a circle represent hyperfine
           interaction.}
  \label{hfs_diagrams}
\end{figure}
\end{center}
%%%%%%%%%%%%%%%%%%%%%%%%%%%
%%%%%%%%%%%%%%%%%%%%%%%%%%%
The measured value of an atomic property $A$ for the atomic state 
$|\Psi_v\rangle $ is the expectation 
\begin{equation}  
  \langle A \rangle = \frac{\langle \Psi_v|A|\Psi_v \rangle}
                      {\langle \Psi_v |\Psi_v \rangle}.
  \label{a_expect}
\end{equation} 
In the present case, $A$ is the hyperfine interaction $H_{\rm hfs}$ and in 
particular the magnetic dipole hyperfine interaction. From here on
we use $H_{\rm hfs}$, however, the derivations and discussions are general,
applicable to any dynamical variable. When coupled-cluster wave functions, 
from Eq.(\ref{ccwaveopen}), are chosen as the correlated atomic states
\begin{equation}
  \langle \Psi_v|H_{\rm hfs}|\Psi_v \rangle = 
       \langle \Phi_v| \tilde H_{\rm hfs} + 2 S^\dagger \tilde H_{\rm hfs} + 
       S^\dagger \tilde H_{\rm hfs} S |\Phi_v\rangle ,
  \label{hfs_num}
\end{equation}
where, $\tilde H_{\rm hfs} = e{^T}^\dagger H_{\rm hfs} e^T$ is the dressed 
operator. The factor of two in the second term on the right hand side accounts 
for $\tilde H_{\rm hfs} S $ as  
$S^\dagger \tilde H_{\rm hfs} = \tilde H_{\rm hfs} S $. An expansion of 
$\tilde H_{\rm hfs}$ ideal for an order wise calculation is
\begin{equation}
  \tilde H_{\rm hfs} = H_{\rm hfs} e^T + \sum_{n = 1}^\infty \frac{1}{n!}
                 \left ( T^\dagger \right )^n H_{\rm hfs} e^T.
  \label{A_tilde}
\end{equation}
The normalization factor, denominator in Eq.(\ref{a_expect}), in terms of 
coupled-cluster wave function is
\begin{equation}
  \langle \Psi_v |\Psi_v \rangle = 
      \langle \Phi_v|\left (1 + S^\dagger\right ) e{^T}^\dagger e^T
                     \left ( 1 +  S\right )|\Phi_v\rangle.
\end{equation}
The dressed operator $ \tilde H_{\rm hfs}$ and operator $ e{^T}^\dagger e^T$
in the normalization factor are non terminating series. In the next section
we describe a method to calculate $ \tilde H_{\rm hfs}$ to all order in $T$ 
iteratively. To our knowledge, this is the first ever implementation of 
such a method within coupled-cluster theory.

%%%%%%%%%%%%%%%%%%%%%%%%%%%%%%%%%%%%%%%%%%%%%%%%%%%%%%%%%%%%%%%%%%%%%%%%%%%%%%%%
%%%%%               Section: Properties to all order                       %%%%%
%%%%%%%%%%%%%%%%%%%%%%%%%%%%%%%%%%%%%%%%%%%%%%%%%%%%%%%%%%%%%%%%%%%%%%%%%%%%%%%%

\section{Properties to all order} 
  \label{all-order}

\begin{figure}[h]
  \includegraphics[width = 8.5cm]{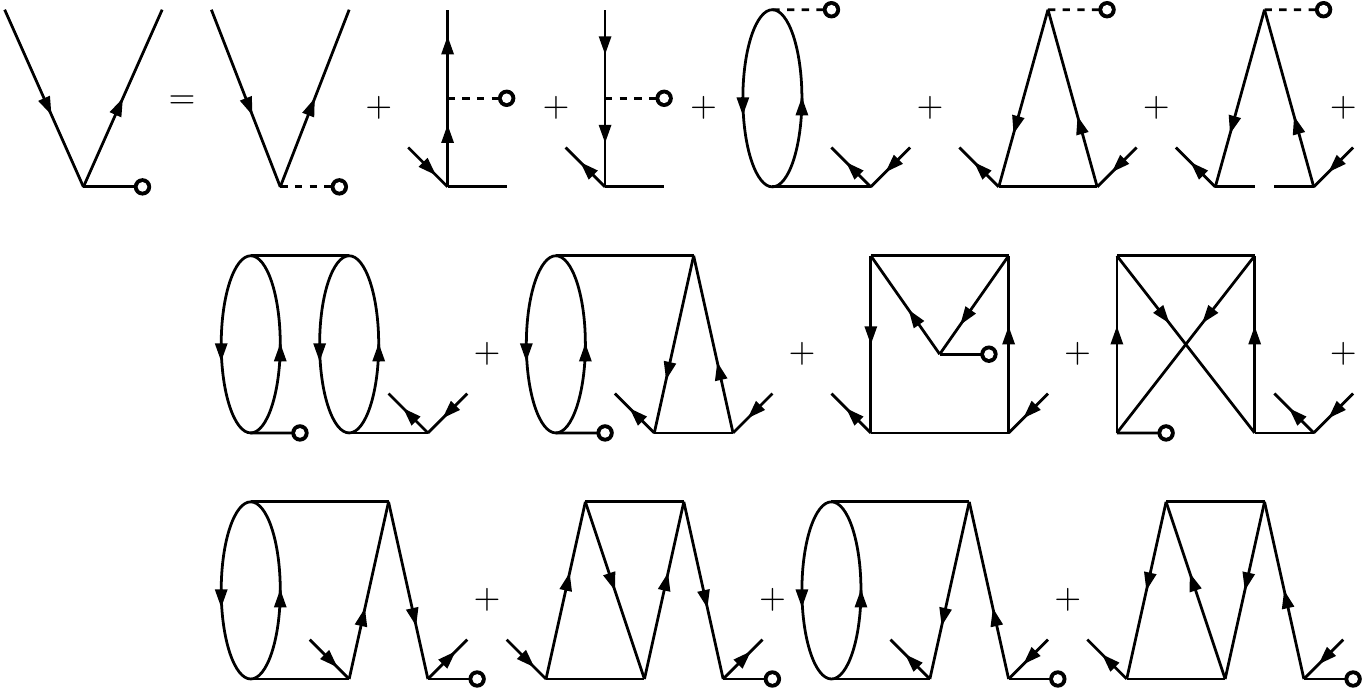}
  \caption{Diagrammatic representation of the iterative equation to calculate 
           the {\em loe} one effective hyperfine operator 
           ${H_{\rm hfs}}_1^{\rm eff}$.  The iteration is implemented with the 
           $T_2^\dagger$ and $T_2$.}
  \label{all_order_eff}
\end{figure}

 For accurate properties calculations it is appropriate to include higher
order terms in $\tilde H_{\rm hfs}$. It is however non trivial to go beyond the
second order, the number of diagrams is large and a systematic evaluation
is extremely tedious. On the other hand, diagrams can be grouped into
different level of excitation ({\em loe}) and evaluate order wise iteratively. 
Here, {\em loe} is the number of core or valence electrons replaced with 
virtual electrons. For example, the diagrams in Fig.\ref{all_order_eff} 
have {\em loe} one. In each of these diagrams, one core electron is replaced 
by a virtual electron.

 To calculate the diagrams of {\em loe} one to all order, consider 
the {\em loe} one diagrams arising from $H_{\rm hfs}e^T$. That is 
\begin{equation}
   \left ( H_{\rm hfs}e^T\right )_1 =  \left ( H_{\rm hfs} + 
         \contraction[0.5ex]{}{H_{\rm hfs}}{}{T}H_{\rm hfs}T + 
         \frac{1}{2}\contraction[0.5ex]{}{H_{\rm hfs}}{}{T}
         \contraction[0.8ex]{}{H_{\rm hfs}}{T}{T}H_{\rm hfs}TT \right )_1 .
   \label{iter_zero}
\end{equation}
Where the subscript denotes the {\em loe} of the contributing terms. It is
equivalent to a one-particle interaction and considered as effective 
properties operator which incorporates electron correlations. In the next 
iteration 
\begin{eqnarray}
   \left ( T^\dagger H_{\rm hfs}e^T\right )_1 &=&  \sum_i\left [ T^\dagger_i 
         \left ( H_{\rm hfs} + 
         \frac{1}{2}\contraction[0.5ex]{}{H_{\rm hfs}}{}{T}H_{\rm hfs}T 
         \right .  \right .     \nonumber  \\
         && \left . \left . + 
         \frac{1}{6} \contraction[0.5ex]{}{H_{\rm hfs}}{}{T} 
         \contraction[0.8ex]{}{H_{\rm hfs}}{T}{T}H_{\rm hfs}TT 
         \right ) T_i \right ]_1^{\rm conn} ,
\end{eqnarray}
where $i=1,2$ in CCSD and the superscript {\em conn} imply only the connected
diagrams contribute. From the definition of the cluster operators,
$T_i$ and $T^\dagger _i$ have {\em loe} $i$ and $-i$ respectively. The above 
equation is equivalent to the expression in Eq.(\ref{iter_zero}) sandwiched
between cluster operators of equal but opposite {\em loe}. So the net {\em loe}
remains unchanged. In general, we can then write
\begin{equation}
   \left ( { T^\dagger } ^n H_{\rm hfs}e^T\right )_1 =  
         \sum_i\left [ T^\dagger_i \left ( { T^\dagger } ^{n-1} 
         H_{\rm hfs}e^T \right ) _1 T_i \right ]_1^{\rm conn} .
\end{equation}
This is an iterative equation and it is possible to evaluate it order by order
to convergence. The sum of all the contributions is equivalent to 
calculating the effective operator
\begin{equation}
  {\cal H}_1= ({e^T}^\dagger H_{\rm hfs}e^T)_1.
\end{equation}
This contribute to the hyperfine structure as $S_2^\dagger {\cal H}_1$. At the 
lowest level there are diagrams and correspond to Fig.\ref{hfs_diagrams}j-k. 
In a similar same way, the effective properties of higher {\em loe} are 
calculated.

 For further study, we resort to diagrammatic analysis. Consider diagrams 
arising from $(H_{\rm hfs}e^T)_1$, there are six diagrams in total. 
These are shown on the first row at the right hand side of 
Fig.\ref{all_order_eff}. These define the initial choice of the effective 
diagram. For the next and higher iterations, consider the contractions with 
$T_2^\dagger$ and $T_2$. The contribution from the $T_1^\dagger$ and $T_1$ is 
neglected as these cluster amplitudes, on an average, are several orders of 
magnitude smaller than $T_2$. Then the iteration is equivalent to the 
diagrammatic equation in Fig.\ref{all_order_eff} and it is mathematically
\begin{equation}
  {\cal H}_1=  {\cal H}_1^0 +  \left ( T_2^\dagger {\cal H}_1T_2\right )_1. 
  \label{iter_eqn}
\end{equation}
Where ${\cal H}_1^0 $ is $(H_{\rm hfs}e^T)_1$, the effective operator prior
to the iteration. Since only the unique diagrams are considered, there are no 
multiplying factors. The algebraic relation in Eq.(\ref{iter_eqn}) is also 
without multiplying factors as the sequence of the contraction is uniquely
defined. Which is not the case in the expansion of 
$e{^T}^\dagger H_{\rm hfs} e^T $.

%%%%%%%%%%%%%%%%%%%%%%%%%%%%%%%%%%%%%%%%%%%%%%%%%%%%%%%%%%%%%%%%%%%%%%%%%%%%%%%
%%%%%            Section: Results and discussions                        %%%%%%
%%%%%%%%%%%%%%%%%%%%%%%%%%%%%%%%%%%%%%%%%%%%%%%%%%%%%%%%%%%%%%%%%%%%%%%%%%%%%%%

\section{Description of numerical methods}
  The calculations presented in the paper involve various numerical techniques
and methods. Some are fairly straight forward, oft used in atomic theory 
calculations. Others are not, specialized and application specific. For 
easy reference in future works, we provide an outline of the numerical methods 
used. This is appropriate as we recommend, based on the current work, an 
approximation of the properties operator in coupled-cluster theory.

%%%%%%%%%%%%%%%%%%%%%%%%%%%%%%%%%%%%%%%%%%%%%%%%%%%%%%%%%%%%%%%%%%%%%%%%%%%%%%%
%%%%%           Subsection: Description of numerical methods             %%%%%%
%%%%%%%%%%%%%%%%%%%%%%%%%%%%%%%%%%%%%%%%%%%%%%%%%%%%%%%%%%%%%%%%%%%%%%%%%%%%%%%

\subsection{Atomic Hamiltonian and single particle states}

 In the results presented in this paper the Dirac-Coulomb Hamiltonian 
is chosen $H^{\rm DC}$ for the calculations. It incorporates relativity at
the single particle level accurately. And, as the name indicates, the 
Coulomb interactions between the electrons. For an $N$ electron atom
\begin{equation}
  H^{\rm DC}=\sum_{i=1}^N\left [c\bm{\alpha}_i\cdot \bm{p}_i+
             (\beta-1)c^2 - V_N(r_i)\right ] +\sum_{i<j}\frac{1}{r_{ij}},
  \label{dchamil}
\end{equation}
where ${\bf p}$ is the linear momentum, and  $\alpha_i$ and $\beta$ are the 
Dirac matrices. For the nuclear potential $V_N(r)$, we consider the  finite 
size Fermi density distribution
\begin{equation}
  \rho_{\rm nuc}(r) = \frac{\rho_0}{1 + e^{(r-c)/a} },
\end{equation}
here, $a = t 4\ln 3$. The parameter $c$ is the half-charge radius, that is
$\rho_{\rm nuc}(c)=\rho_0/2$ and $t$ is the skin thickness. At the single 
particle level, the spin orbitals are of the form
\begin{equation}
  \psi_{n\kappa m}(\bm{r})=\frac{1}{r}
  \left(\begin{array}{r}
            P_{n\kappa}(r)\chi_{\kappa m}(\bm{r}/r)\\
           iQ_{n\kappa}(r)\chi_{-\kappa m}(\bm{r}/r)
       \end{array}\right),
  \label{spin-orbital}
\end{equation}
where $P_{n\kappa}(r)$ and $Q_{n\kappa}(r)$ are the large and small component
radial wave functions, $\kappa$ is the relativistic total angular momentum
quantum number and $\chi_{\kappa m}(\bm{r}/r)$ are the spin or spherical
harmonics. One representation of the radial components is to define these
as linear combination of Gaussian like functions and are referred to as
Gaussian type orbitals (GTOs). Then, the large and small
components \cite{Mohanty-89,Chaudhuri-99} are
\begin{eqnarray}
   P_{n\kappa}(r) = \sum_p C^L_{\kappa p} g^L_{\kappa p}(r),  \nonumber \\
   Q_{n\kappa}(r) = \sum_p C^S_{\kappa p} g^S_{\kappa p}(r).
\end{eqnarray}
The index $p$ varies over the number of the basis functions.
For large component we choose
\begin{equation}
  g^L_{\kappa p}(r) = C^L_{\kappa i} r^{n_\kappa} e^{-\alpha_p r^2},
\end{equation}
here $n_\kappa$ is an integer. Similarly, the small component are
derived from the large components using kinetic balance condition. The
exponents in the above expression follow the general relation
\begin{equation}
  \alpha_p = \alpha_0 \beta^{p-1}.
  \label{param_gto}
\end{equation}
The parameters $\alpha_0$ and $\beta$ are optimized for each of the ions to 
provide good description of the properties. In our case the optimization is to
reproduce the numerical result of the total and orbital energies. The
optimized parameters used in the calculations are listed in 
Table.\ref{tab:basis-param}.

   From Eq.(\ref{spin-orbital}) the reduced matrix element of the magnetic 
hyperfine operator between two spin orbitals , $v'$ and $v$, is
\begin{eqnarray}
  \langle v'||t^1||v\rangle &=& -(\kappa_v + \kappa_{v'})
  \langle -\kappa_{v'}||C^1||\kappa_v \rangle \times \nonumber \\
  &&\int^\infty_0 \frac{dr}{r^2}(P_{n_{v'}\kappa_{v'}} Q_{n_v\kappa_v}
                                  + Q_{n_{v'}\kappa_{v'}} P_{n_v\kappa_v}).
  \label{hfs_matrix}
\end{eqnarray}
A detailed derivation is given in Ref. \cite{Johnson-07}.

%%%%%%%%%%%%%%%%%%%%%%%%%%%%%%%%%%%%%%%%%%%%%%%%%%%%%%%%%%%%%%%%%%%%%%%%%%%%%%
%%%%       Subsection:Basis set and cluster amplitudes                  %%%%%%
%%%%%%%%%%%%%%%%%%%%%%%%%%%%%%%%%%%%%%%%%%%%%%%%%%%%%%%%%%%%%%%%%%%%%%%%%%%%%%

\subsection{Basis set and cluster amplitudes}

%%%%%%%%%%%%%%%%%%%%%%%%%%%%%%%%%%%%%%%%
%%  Basis parameters, alpha and beeta 
%%%%%%%%%%%%%%%%%%%%%%%%%%%%%%%%%%%%%%%%
\begin{table*}[h]
\caption{Optimized parameters $\alpha$ and $\beta$ of the GTO basis used in 
         the calculations. }
  \label{tab:basis-param}
\begin{ruledtabular}
\begin{tabular}{ccccccccccccc}
Symmetry &\multicolumn{3}{c}{$^{25}$Mg$^+$}&\multicolumn{3}{c}{$^{43}$Ca$^+$}
         &\multicolumn{3}{c}{$^{87}$Sr$^+$}&\multicolumn{3}{c}{$^{137}$Ba$^+$}\\
 \hline
    &$\alpha$&$\beta$&Basis&$\alpha$&$\beta$& Basis &
     $\alpha$&$\beta$&Basis&$\alpha$&$\beta$& Basis      \\
    &        &       &function&   &        & function&
             &       &function&   &        & function    \\
 \hline
$s$ &0.0083&2.8900&28&0.0063&2.8800&29&0.0083&2.9800&30&0.0063&2.9800&31 \\ 
$p$ &0.0072&2.9650&25&0.0072&2.9650&26&0.0072&2.9650&27&0.0072&2.9590&28 \\
$d$ &0.0070&2.7200&22&0.0070&2.7000&24&0.0070&2.8000&25&0.0070&2.4500&26 \\
\end{tabular}
\end{ruledtabular}
\end{table*}

  For all the alkaline Earth metal ions considered, Mg$^+$, Ca$^+$, Sr$^+$
and Ba$^+$, we use $V^{N-2}$ orbitals for the calculations. This is equivalent 
to calculating the spin orbitals from the single particle eigenvalue equations 
of the doubly ionized alkaline Earth metal atoms, namely Mg$^{2+}$, Ca$^{2+}$, 
Sr$^{2+}$ and Ba$^{2+}$. Then the single particle basis sets have few bound 
states and rest are continuum. We optimize the basis such that: single 
particle energies of the core and valence orbitals are in good agreement with
the numerical results. For this we use GRASP92 \cite{Parpia-96} to generate
the numerical results.

 As mentioned in earlier sections, we compute the closed-shell cluster 
amplitudes $T$ first. These are used to generate the open shell cluster 
amplitudes $S$. The coupled nonlinear and linear equations are solved 
iteratively. We employ direct inversion in the iterated subspace (DIIS)
\cite{Pulay-80} for convergence acceleration.

%%%%%%%%%%%%%%%%%%%%%%%%%%%%%%%%%%%%%%%%%%%%%%%%%%%%%%%%%%%%%%%%%%%%%%%%%%%%%%%
%%%%%            Section: Results and discussions                        %%%%%%
%%%%%%%%%%%%%%%%%%%%%%%%%%%%%%%%%%%%%%%%%%%%%%%%%%%%%%%%%%%%%%%%%%%%%%%%%%%%%%%

\section{Results and discussions}

%%%%%%%%%%%%%%%%%%%%%%%%%%%%%%%%%%%%%%%%%%%%%%%%%%%%%%%%%%%%%%%%%%%%%%%%%%%%%%
%%%%       Subsection:Attachement and excitation energies               %%%%%%
%%%%%%%%%%%%%%%%%%%%%%%%%%%%%%%%%%%%%%%%%%%%%%%%%%%%%%%%%%%%%%%%%%%%%%%%%%%%%%

\subsection{Ionization potential and excitation energies}

\begin{table*}[t]
\caption{Ionization potential and excitation energies. For comparison
         other results and experimental values are also listed.
                           All values are in atomic units.}
\label{tab-ip-ee}
\begin{ruledtabular}
\begin{tabular}{cccccccccc}
Ion& state &\multicolumn{2}{c}{MBPT}&\multicolumn{2}{c}{Coupled-cluster}
                          &\multicolumn{2}{c}{Other works}
                          &Exp results {Ref\cite{nist}.}                    \\
\hline                                                                      \\
   &       & IP & EE & IP & EE & IP & EE & EE                               \\
\hline                                                                      \\
$^{25}$Mg$^+$&3$s_{1/2}$&-0.55156&0.0    &-0.55203&0.0    &-0.55252 &0.0&0.0\\
      &3$d_{3/2}$&-0.22652&0.32504&-0.22666&0.32537&-0.22677 &0.32575
                                                   \footnotemark[1] &0.32573\\
      &3$d_{5/2}$&-0.22652&0.32504&-0.22668&0.32535&-0.22677 &0.32575
                                                   \footnotemark[1] &0.32574\\
      &3$p_{1/2}$&-0.38922&0.16234&-0.38950&0.16253&-0.39003 &0.16249
                                                   \footnotemark[1] &0.16252\\
      &3$p_{3/2}$&-0.38878&0.16278&-0.38917&0.16286&-0.38961 &0.16291
                                                   \footnotemark[1]&0.16294 \\
                      \\                         
$^{43}$Ca$^+$&4$s_{1/2}$&-0.43784&0.0    &-0.43671&0.0    &-0.43836&0.0&0.0 \\
      &3$d_{3/2}$&-0.37797&0.05987&-0.37601&0.06070&-0.37768&0.06068
                                                   \footnotemark[2]&0.06220 \\
      &3$d_{5/2}$&-0.37762&0.06022&-0.37578&0.06093&-0.37731&0.06205
                                                   \footnotemark[2]&0.06247 \\
      &4$p_{1/2}$&-0.32180&0.11604&-0.32128&0.11543&-0.32217&0.11619
                                                   \footnotemark[2]&0.11478 \\
      &4$p_{3/2}$&-0.32075&0.11709&-0.32119&0.11552&-0.32111&0.11725
                                                   \footnotemark[2]&0.11580 \\
                      \\                         
$^{87}$Sr$^+$&5$s_{1/2}$&-0.40788&0.0    &-0.40573&0.0    &-0.40839&0.0&0.0 \\
      &4$d_{3/2}$&-0.34236&0.06552&-0.33926&0.06647&-0.34279&0.06560
                                                   \footnotemark[2]&0.06632\\
      &4$d_{5/2}$&-0.34091&0.06697&-0.33827&0.06746&-0.34132&0.06707
                                                   \footnotemark[2]&0.06760\\
      &5$p_{1/2}$&-0.29793&0.10995&-0.29696&0.10877&-0.29838&0.11001
                                                   \footnotemark[2]&0.10805\\
      &5$p_{3/2}$&-0.29421&0.11367&-0.29425&0.11148&-0.29463&0.11376
                                                   \footnotemark[2]&0.11171\\
                      \\                         
$^{137}$Ba$^+$&6$s_{1/2}$&-0.37297&0.0    &-0.36862&0.0   &-0.37308&0.0&0.0\\
      &5$d_{3/2}$&-0.35296&0.02001&-0.34758&0.02104&-0.35172&0.02136
                                                   \footnotemark[2]&0.02221\\
      &5$d_{5/2}$&-0.34872&0.02425&-0.34386&0.02476&-0.34748&0.02560
                                                   \footnotemark[2]&0.02586\\
      &6$p_{1/2}$&-0.27685&0.09612&-0.27483&0.09379&-0.27532&0.09776
                                                   \footnotemark[2]&0.09232\\
      &6$p_{3/2}$&-0.26882&0.10415&-0.26821&0.10041 &-0.26946&0.10362
                                                   \footnotemark[2]&0.10002
\end{tabular}
\end{ruledtabular}
\footnotetext[1]{Reference\cite{Safronova-98}.}
\footnotetext[2]{Reference\cite{Guet-91}.}
\end{table*}

 To determine the quality of the basis set and parameters, we compute the 
attachment energies of the ground state ($S_{1/2}$) and the first 
excited $P_{1/2}$, $P_{3/2}$, $D_{3/2}$ and $D_{5/2}$ states are calculated. 
Then the ionization potential (IP), the energy required to remove the valence 
electron, is the negative of the attachement energy $ -E^{\rm att}$. 
To calculate the excitation energy (EE) of the state $|\Psi_v\rangle $, 
consider $E^{\rm att}_g$ and $ E^{\rm att}_v$ as the attachment 
energies of the ground state and excited state. Then difference 
$ E^{\rm att}_v- E^{\rm att}_g$ is the EE, it can as well be 
defined in terms of IPs.

  For further analysis on the correlation effects incorporated with CCSD, we 
first compute IP with relativistic many-body perturbation theory (MBPT). The 
MBPT calculations are similar to our previous work \cite{Mani-09} for second 
order correlation energy of closed-shell atoms, in particular noble gas atoms. 
The MBPT diagrams of IP are similar to the first four attachment diagrams in 
Fig.\ref{fig-att-diag} but have residual interaction in stead of $S$ and $T$ 
operators. Where, the first two diagrams Fig.\ref{fig-att-diag}(a-b), direct 
and exchange, have the valence replaced by a virtual state and encapsulates 
core-valence correlation. The remaining diagrams Fig.\ref{fig-att-diag}(c-d)
represent core-core correlation as these involve double replacement of core 
electrons. The other two diagrams with $T_1$ do not contribute as single 
replacements with residual Coulomb interaction are zero.

 The results of the MBPT calculations are listed in Table.\ref{tab-ip-ee}. 
The $3d\; ^2D_{3/2, 5/2}$ and $3p\; ^2P_{1/2, 3/2}$ of Mg$^+$, evaluated from 
the MBPT IPs, are marginally lower than the experimental data but are very 
close. From Ca$^+$, there is a change in the pattern of the EEs. The 
MBPT results of $^2D_{3/2, 5/2}$ EEs are lower than the experimental data,
whereas the $ ^2P_{1/2, 3/2}$ EEs are higher. The same pattern occurs in 
Sr$^+$ and Ba$^+$. Similar pattern is observed in the results of previous 
calculations \cite{Guet-91}. The differences between the results 
in Ref.\cite{Guet-91} and ours are minor  and random in nature. These 
deviations can be attributed to the nature and completeness of the 
basis sets chosen in the two calculations. 

  The CCSD results of the EE are also listed in Table.\ref{tab-ip-ee}, these
are closer to the experimental data than the MPBT results. This is not 
surprising as CCSD encapsulates electron correlations more accurately. The
trend of the CCSD results separates into two: Mg$^+$ and other ions. The 
additional electron correlation increases the IPs of Mg$^+$, whereas there
is a decrease in the IPs of Ca$^+$, Sr$^+$ and Ba$^+$. However, the states 
change differently such that the EE improves. These results vouch for the 
reliability of the basis set for properties calculations.

%%%%%%%%%%%%%%%%%%%%%%%%%%%%%%%%%%%%%%%%%%%%%%%%%%%%%%%%%%%%%%%%%%%%%%%%%%%%%%
%%%%       Subsection:Magnetic dipole hyperfine constants               %%%%%%
%%%%%%%%%%%%%%%%%%%%%%%%%%%%%%%%%%%%%%%%%%%%%%%%%%%%%%%%%%%%%%%%%%%%%%%%%%%%%%

\subsection{Magnetic dipole hyperfine constants}

\begin{table*}
\caption{Magnetic dipole hyperfine structure constants (in MHz) for 
         $^{25}$Mg$^+$, $^{43}$Ca$^+$, $^{87}$Sr$^+$, and $^{137}$Ba$^+$ ions.}
\label{tab-hfs}
\begin{ruledtabular}
\begin{tabular}{ccccc}
Ion & state & This work & Other works & Experiment   \\
\hline
$^{25}$Mg$^+$&$3s_{1/2}$&$-596.785$&$-597.6$\footnotemark[12],
                                   $-554$\footnotemark[19],
                                   $-(602\pm8)$\footnotemark[20]
                                  &$-596.254$\footnotemark[13]               \\
      &$3p_{1/2}$&$-102.997$&$-103.4$\footnotemark[12],$-100$\footnotemark[19]
                                                                          &- \\
      &$3p_{3/2}$&$-19.546$ &$-19.29$\footnotemark[12],$-19$\footnotemark[19]
                                                                          &- \\
      &$3d_{3/2}$&$-1.083$  &$-1.140$\footnotemark[12],$-1.25$\footnotemark[19]
                                                                          &- \\
      &$3d_{5/2}$&$0.118$   &$0.1196$\footnotemark[12],
                            $0.107$\footnotemark[1],$0.17$\footnotemark[19]
                                                                          &- \\
                     \\
$^{43}$Ca$^+$&$4s_{1/2}$&$-808.126$&$-805.35$\footnotemark[2],
                         $-819$\footnotemark[7],
                         $-794.7$\footnotemark[8],$-806.4(2.5)$\footnotemark[21]
                                                 &$-797.5(2.4)$\footnotemark[3]
                                                  $-805(2)$\footnotemark[4]  \\
      &$4p_{1/2}$&$-142.782$&$-143.07$\footnotemark[2],$-148$\footnotemark[7],
                            $-144.8$\footnotemark[8],$-143$\footnotemark[19],
                            $-145.4(4)\footnotemark[21]$
                                                 &$-158(3.3)$\footnotemark[3],
                                                 $-145.5(1.0)$\footnotemark[4],
                                                 $-142(8)$\footnotemark[5],
                                                 $-145.4(0.1)$\footnotemark[6]\\
      &$4p_{3/2}$&$-32.185$ &$-30.50$\footnotemark[2],$-30.9$\footnotemark[7],
                            $-29.3$\footnotemark[8],$-30$\footnotemark[19],
                            $-30.4(4)$\footnotemark[21]
                                                 &$-29.7(1.6)$\footnotemark[3],
                                                 $-31.9(0.2)$\footnotemark[4],
                                                 $-31.0(0.2)$\footnotemark[6]\\
      &$3d_{3/2}$&$-45.294$ &$-47.82$\footnotemark[2],
                           $-52$\footnotemark[7],$-49.4$\footnotemark[8],
                           $-47.3(3)$\footnotemark[21]
                                                 &$-48.3(1.6)$\footnotemark[5],
                                                 $-47.3(0.2)$\footnotemark[6]\\
      &$3d_{5/2}$&$-4.008$  &$-3.351$\footnotemark[1],$-3.55$\footnotemark[2],
                            $-5.2$\footnotemark[7], $-4.2$\footnotemark[8],
                            $-3.6(3)$\footnotemark[21]
                                                &$-3.8(0.6)$\footnotemark[6],
                                                 $3.8931(2)$\footnotemark[22]\\ 
                      \\
$^{87}$Sr$^+$&$5s_{1/2}$&$-990.638$&$-10003.18$\footnotemark[2],
                            $-1000$\footnotemark[11]
                                               &$-1000.5(1.0)$\footnotemark[9]\\
      &$5p_{1/2}$&$-169.988$&$-178.40$\footnotemark[2],$-177$\footnotemark[11],
                            $-175$\footnotemark[19]&-\\
      &$5p_{3/2}$&$-36.225$ &$-35.11$\footnotemark[2],$-35.3$\footnotemark[11]
                            $-30$\footnotemark[19] &$-36.0$\footnotemark[9]\\
      &$4d_{3/2}$&$-44.320$ &$-47.36$\footnotemark[2],
                            $-46.7$\footnotemark[11]&-\\
      &$4d_{5/2}$&$2.168$   &$2.156$\footnotemark[1],
                            $2.51$\footnotemark[2],
                            $1.1$\footnotemark[11]&$2.17$\footnotemark[10]\\
                     \\
$^{137}$Ba$^+$&$6s_{1/2}$&$4021.721$&$4072.83$\footnotemark[16]
                                                    &$4018.2$\footnotemark[17]\\
      &$6p_{1/2}$&$705.039$ &$736.98$\footnotemark[16]&-\\
      &$6p_{3/2}$&$130.191$ &$130.94$\footnotemark[16],$126$\footnotemark[19]
                                                   &$126.9$\footnotemark[15],
                                                    $112.77$\footnotemark[18]\\
      &$5d_{3/2}$&$185.013$ &$192.99$\footnotemark[14],$188.76$\footnotemark[16]
                            $215$\footnotemark[15]&$189.730$\footnotemark[15],
                                                    $170.88$\footnotemark[18]\\
      &$5d_{5/2}$&$-12.593$ &$9.39$\footnotemark[14],$-11.717$\footnotemark[1],
                            $-18$\footnotemark[15]&$-12.028$\footnotemark[15]\\
\end{tabular}
\end{ruledtabular}
\footnotetext[1]{Reference\cite{Sahoo-07}.}
\footnotetext[2]{Reference\cite{Yu-04}.}
\footnotetext[3]{Reference\cite{Goble-90}.}
\footnotetext[4]{Reference\cite{Silverans-91}.}
\footnotetext[5]{Reference\cite{Kurth-91}.}
\footnotetext[6]{Reference\cite{Nortershauer-88}.}
\footnotetext[7]{Reference\cite{Martensson-84}.}
\footnotetext[8]{Reference\cite{Martensson-92}.}
\footnotetext[9]{Reference\cite{Buchinger-90}.}
\footnotetext[10]{Reference\cite{Barwood-03}.}
\footnotetext[11]{Reference\cite{Martensson-02}.}
\footnotetext[12]{Reference\cite{Safronova-98}.}
\footnotetext[13]{Reference\cite{Itano-81}.}
\footnotetext[14]{Reference\cite{Itano-06}.}
\footnotetext[15]{Reference\cite{Silverans-86}.}
\footnotetext[16]{Reference\cite{Sahoo-03}.}
\footnotetext[17]{Reference\cite{Becker-68}.}
\footnotetext[18]{Reference\cite{Hohle-78}.}
\footnotetext[19]{Reference\cite{Heully-85}.}
\footnotetext[20]{Reference\cite{Ahmad-83}.}
\footnotetext[21]{Reference\cite{sahoo-09}.}
\footnotetext[22]{Reference\cite{Benhelm-07}.}
\end{table*}

 To compute the hyperfine constants from the CCSD wave functions, we use
Eq.(\ref{hfs_num}). The results are listed in Table.\ref{tab-hfs}, for 
comparison the results of other theoretical calculations and experimental
data are also given. As defined in Eq.(\ref{hfs_num}), the coupled-cluster 
expression of the hyperfine structure constants is separated into three 
groups. The dominant contribution from the first term $\tilde H_{\rm hfs}$, 
up to first order in $T^\dagger$ and $T$, is 
\begin{equation}
  \tilde H_{\rm hfs} \approx H_{\rm hfs} + 2H_{\rm hfs}T_1 + 
     T_1^\dagger H_{\rm hfs}\left ( T_1 + 2T_2 \right ) + 
     T_2^\dagger H_{\rm hfs}T_2 .
\end{equation}
Here,  the first term is the Dirac-Fock (DF), which has the largest 
contribution. The factor two in the second and fourth terms accounts for the 
complex conjugate terms. The third term, second order in $T_1$, has one 
diagram and negligibly small contribution. The diagrams arising from the last 
term are topologically similar to the attachment diagrams (c-d) in 
Fig.\ref{fig-att-diag}. However, with the $T_2^\dagger$ in stead of residual 
Coulomb interaction  and $ H_{\rm hfs}$ inserted on one of the orbital lines. 
There are ten diagrams and contribution from these are labelled as 
$\tilde H_{\rm hfs}-{\rm DF}$. The last two terms in Eq.(\ref{hfs_num}) are 
approximated as
\begin{eqnarray}
  S^\dagger \tilde H_{\rm hfs}   & \approx &  2S^\dagger  
              \left ( H_{\rm hfs}e^T\right )_1 ,  \\ 
  S^\dagger \tilde H_{\rm hfs} S & \approx &    S_1^\dagger H_{\rm hfs}\left 
              ( S_1 + 2S_2 \right ) + 
     S_2^\dagger H_{\rm hfs}S_2 .
\end{eqnarray}
Like in $\tilde H_{\rm hfs} $, the factor of two is to account for the complex
conjugate terms. The expression of $ ( H_{\rm hfs}e^T)_1$ is as given in 
Eq.(\ref{iter_zero}). The  $S_2^\dagger H_{\rm hfs}S_2$ term have contributions
from the diagrams (b-g) in Fig.\ref{hfs_diagrams}. These are topologically 
similar to (a-b) in Fig.\ref{fig-att-diag}. But, like in 
$ T_2^\dagger H_{\rm hfs}T_2$, $S_2^\dagger$ instead of residual Coulomb 
interaction and $H_{\rm hfs}$ inserted to one orbital line. Diagrams arising
from the remaining terms are also given in Fig.\ref{hfs_diagrams}. Based on
this grouping, the contributions are listed in Table.\ref{tab-hfs-comp}. In the
following we present a detailed comparison of  our magnetic hyperfine 
constants results with the earlier ones. As discussed later, some of our 
results are the best match with experimental data. This is a thorough
test for the starting point of our iterative procedure and the expression
for properties calculation we recommend.

%%%%%%%%%%%%%%%%%%%%%%%%%%%%%%%%%%%%%%%%%%%%%%%%%%%%%%%%%%%%%%%%%%%%%%%%%%%%%%
%%%%       Subsubsection:Description of results for each ion            %%%%%%
%%%%%%%%%%%%%%%%%%%%%%%%%%%%%%%%%%%%%%%%%%%%%%%%%%%%%%%%%%%%%%%%%%%%%%%%%%%%%%

\subsubsection{{\rm Mg}$^+$}

   The experimental data is available only for the ground state 
$3s\; ^2S_{1/2}$ \cite{Itano-81}. However, theoretical results are available
for the low lying states $3s\; ^2S_{1/2}$, $3p\; ^2P_{1/2}$, 
$3p\; ^2P_{3/2}$, $3d\; ^2D_{3/2}$ and $3d\; ^2D_{5/2}$. In the previous 
works, the calculations used relativistic many-body perturbation theory
\cite{Ahmad-83,Heully-85} and linearized CCSD \cite{Safronova-98} using 
numerical and B-splines basis sets respectively. These report the  DF 
contribution for $3s\; ^2S_{1/2}$ as $-466.4$ \cite{Ahmad-83} and  
$-463$ \cite{Heully-85}. The later is in excellent agreement with our result 
$-463.29$. The other dominant terms are $S^\dagger \tilde H_{\rm hfs}$ and 
$\tilde H_{\rm hfs}-{\rm DF}$, contribution from these are $-107.32$ and 
$-16.13$ respectively. Total value of these three terms is $-586.76$, 98\% of 
the experimental value. Our total value $-596.78$, after including 
$S^\dagger H_{\rm hfs}S$, is 0.08\% lower than the experimental value and is 
the best theoretical result.

   For the $3p\; ^2P_{1/2}$ and $3p\; ^2P_{3/2}$ states, our DF 
values $-76.98$ and $-15.24$ are in very good agreement with the values
 $-77$ and $-15.2$ given in Ref. \cite{Heully-85}. The two states have 
20.6\% and 16.3\% contribution to the total value from 
$S^\dagger \tilde H_{\rm hfs}$. This difference shows variations in the nature 
of correlation effects, predominantly core-polarization. Our total values for 
the two states are $-103.0$ and $-19.55$, these are in very good agreement 
with the previous results.

  For the $3d ^2D_{3/2}$ and $3d ^2D_{5/2}$ states, our DF values
are $-1.26$ and $-0.54$ respectively. Whereas, the values in a previous work
\cite{Heully-85} are $-1.61$ and $-0.54$. The results of $3d ^2D_{5/2}$ match
perfectly but there is a significant difference in the results of 
$3d ^2D_{3/2}$. Our result of -1.26 is 28\% less in magnitude. The correlation
effects, core-polarization in particular, are markedly different from the other
states. Contribution from $S^\dagger \tilde H_{\rm hfs}$ to $3d ^2D_{3/2}$ is 
0.19 is 15\% in magnitude of Dirac-Fock and opposite in sign. It is even more 
dramatic for $3d ^2D_{5/2}$, it is 0.65, which larger than  Dirac-Fock in 
magnitude and opposite in sign.

  Considering that the calculations in Ref. \cite{Heully-85} incorporates
core-polarization to all orders , we can extract the pair correlation effects.
For the $3s\;^2S_{1/2}$ state, the core-polarization contributes $-91$. 
Subtracting this from our $S^\dagger \tilde H_{\rm hfs}$ result, the pair 
correlation contribution to this term is $-16.32 $. Adding the other terms as 
well, the total contribution from pair correlation is $-43.82$. Which is less 
than the core-polarization but not negligible. For the other states the 
core-polarization contributions are $-18$, $-3.7$ and $0.71$ for 
$3p\;^2P_{1/2}$, $3p\;^2P_{3/2}$ and $3d\;^2D_{5/2}$ respectively. The 
corresponding pair correlation contribution are $-7.86$, $-0.62$ and
$-0.04$. The pair correlation is negligible in last two states and we have not
estimated for $3d\;^2D_{3/2}$ as there is a large difference between our 
DF value and Ref. \cite{Heully-85}.

\subsubsection{{\rm Ca}$^+$}

  This is the most well studied, experimentally and theoretically, 
singly ionized alkaline Earth ion. There is a large variation 
in the experimental results of $4s\;^2S_{1/2}$ and $4p\; ^2P_{1/2}$, and
less in the results of $4p\; ^2P_{3/2}$, $3d\; ^2D_{3/2}$  and 
$3d\; ^2D_{5/2}$ states. On the other hand the theoretical results exhibit 
significant variations for all the states except $4p\; ^2P_{3/2}$. The 
DF values of $4s\;^2S_{1/2}$ reported in previous works are
$-589$ \cite{Heully-85} and $-588.933$ \cite{Yu-04}, these are calculated with
numerical and B-spline basis sets respectively. Our value $-589.09$ is in 
very good agreement with these results. The core-valence correlation from
$S^\dagger \tilde H_{\rm hfs}$ accounts for 22\% of the 
total value. This is much larger than in Mg$^+$ (17\%). On the other hand 
core-core correlation, contribution from $\tilde H_{\rm hfs}-{\rm DF}$, is 
smaller.  Our total value $-808.12$ is marginally higher than the experimental
values but lies between the other theoretical results.

  In previous studies DF values of the $4p\; ^2P_{1/2}$  are
$-102$ \cite{Heully-85} and  $-101.492$ \cite{Yu-04}. Similarly, 
for $4p\; ^2P_{3/2}$ the values are $-19.2$ \cite{Heully-85} and 
$-19.646$ \cite{Yu-04}. These are in very good agreement with our results 
$-101.47$ and $19.65$. Like in $4s\;^2S_{1/2}$ there is an increase,
compared to Mg$^+$, in  $S^\dagger \tilde H_{\rm hfs}$ contribution. It 
accounts for 26\% and 30\% of the total value for the two states. Our total 
values of $4p\; ^2P_{1/2}$ is lower than the other theoretical results. 
Whereas $4p\; ^2P_{3/2}$ exhibits opposite trend.

 For  $3d\; ^2D_{3/2}$, the DF values in the previous studies are
$-33$ \cite{Heully-85}, $-33.206$ \cite{Yu-04} and $-39.12$ \cite{Itano-06}. 
The first two compares well with our value $-33.55$. Similarly, our 
$3d\; ^2D_{5/2}$ DF value $-14.29$ is in good agreement with the previous 
results $-14$ \cite{Heully-85} and $-14.144$ \cite{Yu-04}. There is a change 
in the nature of $S^\dagger \tilde H_{\rm hfs}$ contribution to
$3d\; ^2D_{3/2}$. Unlike in Mg$^+$, it is in phase with DF and similar trend 
is observed in Sr$^+$ and Ba$^+$ as well. The contribution from
$(S^\dagger \tilde H_{\rm hfs} + {\rm c.c.})$ to $3d\; ^2D_{5/2}$ is the 
only one which is less in magnitude than the DF value. In all the other
ions ( Mg$^+$, Sr$^+$ and Ba$^+$) DF values are less in 
magnitude. The impact of core-core correlation is not large but not negligible.
Our total value  for $3d\; ^2D_{3/2}$ is lower than all the theoretical and 
experimental values. However, our result for $3d\; ^2D_{5/2}$ matches very
well with the experimental data.

  Taking the core-polarization results from Ref. \cite{Yu-04} and following
the procedure in Mg$^+$ we estimate the pair correlation effects. We get the
pair correlation contributions as $-108.61$, $-19.37$, $-4.25$, $-11.96$
and $-7.99$ for the $4s\;^2S_{1/2}$ and $4p\; ^2P_{1/2}$, $4p\; ^2P_{3/2}$ 
$3d\; ^2D_{3/2}$ and $3d\; ^2D_{5/2}$ respectively. Except for 
$4p\; ^2P_{3/2}$ and $3d\; ^2D_{3/2}$, these are in very good agreement with 
the pair correlation listed in Ref. \cite{Yu-04}. Not surprisingly, our 
results for these two states deviate from the other theoretical and 
experimental data.

\subsubsection{{\rm Sr}$^+$}

  Experimental data is limited to $5s\; ^2S_{1/2}$, $4p\; ^2P_{1/2}$  and
$4d\; ^2D_{5/2}$. However, several theoretical investigations have examined 
the hyperfine structure of Sr$^+$. The $5s\; ^2S_{1/2}$ DF value  earlier 
works are $-735$ \cite{Martensson-02} and $-736.547$ \cite{Yu-04}.
Our value $-738.204$ is higher than both of the values. There is a large 
contribution from $S^\dagger \tilde H_{\rm hfs}$. It is 22\% of the total 
value and same as $4s\;^2S_{1/2}$ of Ca$^+$. The core-core correlation is less 
significant. Our total result is  lower than the experimental data and other 
theoretical results.

 The DF values of $5p\; ^2P_{1/2}$ from previous works are 
$-122$ \cite{Martensson-02} and $-121.576$ \cite{Yu-04}. And values for
$5p\; ^2P_{3/2}$ are $-21.4$ \cite{Martensson-02} and $-21.331$ \cite{Yu-04}. 
These are less than our values $-122.363$ and $-21.501$. The core-core 
correlation effects is negligibly small, 0.3\% of the total value. Compared 
to Ca$^+$ ($4p\;^2P_{J}$), there is an enhanced role of 
$S^\dagger \tilde H_{\rm hfs}$ in $5p\;^2P_{3/2}$. It accounts for 33\% of 
the total value. Our total value for $5p\; ^2P_{1/2}$ is less than
the previous theoretical results. But, the value of $5p\; ^2P_{1/2}$ is in 
excellent agreement with the experimental data. 

 The DF values of $4d\;^2D_{3/2}$ from previous works are
$-31.2$ \cite{Martensson-02}, $-31.126$ \cite{Yu-04} and $-34.23$
\cite{Itano-06}. And values for $4d\;^2D_{5/2}$  are $-13.0$ 
\cite{Martensson-02}, $-12.977$ \cite{Yu-04} and $-14.27$ \cite{Itano-06}. 
These compare well with our values $-31.368$ and $-13.080$. There is a marked 
change in the role of $S^\dagger \tilde H_{\rm hfs}$ for the $4d\; ^2D_{5/2}$ 
state. It has larger magnitude ( 135\% ) than the DF value. Our total value
for $4d\;^2D_{3/2}$ is lower than the other theoretical values. However,
$4d\;^2D_{3/2}$ is in excellent agreement with the experimental data.

 There is noticeable difference in the estimates of the core-polarization
effects in the earlier works \cite{Martensson-02,Yu-04}. For example, the
core-polarization contribution to $4d\;^2D_{3/2}$ is estimated as
$-6.3$ in Ref. \cite{Martensson-02}, whereas it is $-2.413$ in 
Ref. \cite{Yu-04}. For consistency of analysis, with the choice in Ca$^+$, we 
adopt the core-polarization results of Ref. \cite{Yu-04} and estimate pair
correlation effects in our results. These are $-127.795$, $-24.241$, 
$-3.938$, $-11.235$ and $-6.594$. After accounting for the difference in the
Dirac-Fock results, the results for $5s\; ^2S_{1/2}$ and $4d\; ^2D_{5/2}$
are in very good agreement with Ref. \cite{Yu-04}.

\subsubsection{{\rm Ba}$^+$}

  It is a candidate system, as mentioned earlier, for a novel technique to 
measure parity nonconservation (PNC) experiment \cite{Fortson-93}. In this 
context, theoretical study of Ba$^+$ hyperfine constants is very important. 
It is a good proxy for the PNC in ions or atoms arising from neutral weak 
currents. Except for $6p\;^2P_{1/2}$, there are experimental data for the 
low-lying states and theoretical results are available for $6s\;^2S_{1/2}$, 
$6p\;^2P_{1/2}$, $6p\;^2P_{3/2}$, $5d\;^2D_{3/2}$ and $5d\;^2D_{5/2}$. The
DF value of $6s\;^2S_{1/2}$ in previous calculations are 
$2929.41$  \cite{Sahoo-03} and $3055$ \cite{Ahmad-82}. Our result is 
$3003.105$ and significantly different from both of the values. The 
contribution from the core-core correlation $\tilde H_{\rm hfs}-DF$ is 
of opposite phase to the DF contribution. This is in contrast to the 
states we have discussed so far. The total value is in very good agreement 
with experimental data but significantly different from the other theoretical 
results. It has 23\% contribution from $S^\dagger \tilde H_{\rm hfs}$. 

 The DF value of $6p\;^2P_{1/2}$ and $6p\;^2P_{3/2}$ in the previous 
work are $492.74$ \cite{Sahoo-03} and $71.84$ \cite{Sahoo-03}. These are 
different from our values of $504.196$ and $73.674$.  The core-core 
correlation $\tilde H_{\rm hfs}-{\rm DF}$, as in $6s\;^2S_{1/2}$, is of 
opposite phase for $6p\;^2P_{1/2}$. The total results of the two states are 
$705.039$ and $130.191$. The first is lower than the previous theoretical 
result. And the later is in very good agreement with the theoretical result 
but lower than the experimental data.

 The DF values of $5d\;^2D_{3/2}$ in the previous studies are 
$128.27$ \cite{Sahoo-03} and $139.23$ \cite{Itano-06}. And for $5d\;^2D_{5/2}$
the values are $53.213$ \cite{Sahoo-07} and $55.82$ \cite{Itano-06}. Our 
results are $129.875$ and $52.085$, these are closer to Ref.\cite{Sahoo-03} and
Ref.\cite{Sahoo-07} respectively. The $S^\dagger \tilde H_{\rm hfs}$ 
contribution to $5d\;^2D_{5/2}$ is large, 141\% of the DF value and opposite 
in phase. Our total total values $185.013$ and $-12.592$ are close to 
experimental data.

  For Ba$^+$, except for the ground state there are no systematic studies 
of core-polarization effects.  The previous work of Ref. \cite{Sahoo-03} uses
methods, GTO basis and relativistic coupled-cluster, similar to what we 
have used in the present paper. Comparing the two, there is a good correlation 
between different coupled-cluster terms for all the states except 
$6s\;^2S_{1/2}$.

%%%%%%%%%%%%%%%%%%%%%%%%%%%%%%%%%%%%%%%%%
%%%%%%%%%%%%%%%%%%%%%%%%%%%%%%%%%%%%%%%%%
\begin{table*}[h]
\caption{Contributions from different terms in the coupled-cluster, 
         magnetic dipole hyperfine constant, properties expression. 
         The values listed are in MHz.}
\label{tab-hfs-comp}
\begin{ruledtabular}
\begin{tabular}{cccccccccc}
Ion& state &\multicolumn{7}{c}{Coupled-cluster terms}            \\
\hline                                                           \\
   &       & DF & $\tilde H_{\rm hfs}$-DF & $S^\dagger\tilde H_{\rm hfs}$
           & $S^\dagger_2\tilde H_{\rm hfs} S_1$
           & $S^\dagger_1\tilde H_{\rm hfs} S_1$
           & $S^\dagger_2\tilde H_{\rm hfs} S_2$
           & $S^\dagger_2 H_{\rm hfs} T+$ & Norm               \\
   &       &    &               & $+ c.c$ & $+ c.c.$ & & &  
           $S^\dagger_2 H_{\rm hfs} T_1 S_1$  &           \\ 
   &       &    &       &        &    &    &  &$+c.c.$ &         \\
 \hline                                                          \\
$^{25}$Mg$^+$&3$s_{1/2}$&-463.297&-16.136&-107.325&-1.532&-0.396&
                                                     -5.560&-3.878&1.002   \\
      &3$p_{1/2}$&-76.984 &-2.754 &-21.254&-0.326&-0.089&-0.989&-0.720&
                                                                      1.001\\
      &3$p_{3/2}$&-15.242 &-0.695 &-3.184 &0.005 &-0.018&-0.277&-0.160&
                                                                      1.001\\
      &3$d_{3/2}$&-1.259  &-0.007 &0.186  &0.004 &-0.001&-0.008&-0.000&
                                                                      1.001\\
      &3$d_{5/2}$&-0.540  &-0.003 &0.648  &0.017 &-0.000&-0.004&-0.000&
                                                                      1.001\\
                      \\                         
$^{43}$Ca$^+$&4$s_{1/2}$&-589.087&-12.696&-180.217&-4.313&-1.802&-10.717&
                                                              -16.771&1.009\\
      &4$p_{1/2}$&-101.473&-0.497 &-37.514&-0.978&-0.446&-1.031 &
                                                              -1.638 &1.006\\
      &4$p_{3/2}$&-19.648 &-0.321 &-9.691 &-0.222&-0.094&-1.004 &
                                                              -1.426 &1.007\\
      &3$d_{3/2}$&-33.554 &-2.553 &-7.701 &-0.149&-0.260&-2.153 &
                                                              0.258  &1.018\\
      &3$d_{5/2}$&-14.294 &-1.247 &13.430 &0.481 &-0.111&-2.449 &
                                                              0.110  &1.018\\
                      \\                         
$^{87}$Sr$^+$&5$s_{1/2}$&-738.204&-3.667&-218.305&-5.258&-3.046&-15.027&
                                                              -18.379&1.011\\
      &5$p_{1/2}$&-122.363&-0.675&-44.231 &-1.120&-0.678&-1.446 &
                                                              -0.637 &1.007\\
      &5$p_{3/2}$&-21.501 &-0.398&-12.011 &-0.339&-0.126&-1.043 &
                                                              -1.099 &1.008\\
      &4$d_{3/2}$&-31.368 &-3.084&-8.431  &-0.271&-0.139&-1.979 &
                                                              0.255  &1.016\\
      &4$d_{5/2}$&-13.080 &-1.626&17.644  &0.470 &-0.058&-1.991 &
                                                              0.843  &1.016\\
                      \\                         
$^{137}$Ba$^+$&6$s_{1/2}$&3003.105&-39.093 &939.272&23.989&17.598&
                                                      66.108&68.032&1.014\\
      &6$p_{1/2}$&504.196 &-5.948 &196.073&5.397 &4.072 &6.064 &1.982 &
                                                                      1.010\\
      &6$p_{3/2}$&73.674  &0.665  &45.835 &1.555 &0.619 &4.480 &4.797 &
                                                                      1.011\\
      &5$d_{3/2}$&129.875 &12.565 &37.918 &1.148 &0.462 &9.495 &-2.329&
                                                                      1.022\\
      &5$d_{5/2}$&52.085  &7.240  &-73.611&-1.520&0.191 &9.554 &-6.803&
                                                                      1.022\\
\end{tabular}
\end{ruledtabular}
\end{table*}

%%%%%%%%%%%%%%%%%%%%%%%%%%%%%%%%%%%%%%%%%%%%%%%%%%%%%%%%%%%%%%%%%%%%%%%%%%%%%%
%%%%                Subsection:All order calculations                   %%%%%%
%%%%%%%%%%%%%%%%%%%%%%%%%%%%%%%%%%%%%%%%%%%%%%%%%%%%%%%%%%%%%%%%%%%%%%%%%%%%%%

\subsection{All order calculations}

%%%%%%%%%%%%%%%%%%%%%%%%%%%
\begin{table*}[h]
\caption{Magnetic dipole hyperfine structure constant, contributions from 
         higher-order terms in the all order scheme Eq.(\ref{A_tilde}).}
 \label{tab:all_order}
\begin{ruledtabular}
\begin{tabular}{ccccccc}
Ion& state &\multicolumn{4}{c}{$S^\dagger\tilde H_{\rm hfs}$}              \\
\hline                                                                     \\
   &       & iter = 0 & iter = 1 & iter = 2 & iter = 3 & Converged         \\
   &       & $(H_{\rm hfs}e^T)_1$ & $T^\dagger_2(H_{\rm hfs}e^T)_1 T_2$
           & ${T^\dagger_2}^2(H_{\rm hfs}e^T)_1 T_2^2$
           & ${T^\dagger_2}^3(H_{\rm hfs}e^T)_1 T_2^3 $
           & value                                                         \\
 \hline                                                                    \\
$^{25}$Mg$^+$&3$s_{1/2}$&-53.663&-53.502&-53.503&-53.503&-53.503 \\
      &3$p_{1/2}$&-10.627&-10.563&-10.564&-10.564&-10.564        \\
      &3$p_{3/2}$&-1.592 &-1.577 &-1.577 &-1.577 &-1.577         \\
      &3$d_{3/2}$&0.093  &0.091  &0.091  &0.091  &0.091          \\
      &3$d_{5/2}$&0.324  &0.321  &0.321  &0.321  &0.321          \\
                      \\                         
$^{43}$Ca$^+$&4$s_{1/2}$&-90.109&-89.776&-89.778&-89.778&-89.778 \\
      &4$p_{1/2}$&-18.757&-18.570&-18.574&-18.574&-18.574        \\
      &4$p_{3/2}$&-4.845 &-4.792 &-4.793 &-4.793 &-4.793         \\
      &3$d_{3/2}$&-3.851 &-3.887 &-3.885 &-3.885 &-3.885         \\
      &3$d_{5/2}$&6.715  &6.638  &6.639  &6.639  &6.639          \\
                      \\                         
$^{87}$Sr$^+$&5$s_{1/2}$&-109.153&-108.716&-108.720&-108.720&-108.720\\
      &5$p_{1/2}$&-22.116&-21.908&-21.912&-21.912&-21.912        \\
      &5$p_{3/2}$&-6.006 &-5.943 &-5.944 &-5.944 &-5.944         \\
      &4$d_{3/2}$&-4.216 &-4.267 &-4.265 &-4.265 &-4.265         \\
      &4$d_{5/2}$&8.822  &8.687  &8.689  &8.689  &8.689          \\
                      \\                         
$^{137}$Ba$^+$&6$s_{1/2}$&469.636&467.423&467.450&467.449&467.449\\
      &6$p_{1/2}$&98.036 &97.052 &97.075 &97.074 &97.074         \\
      &6$p_{3/2}$&22.917 &22.655 &22.660 &22.660 &22.660         \\
      &5$d_{3/2}$&18.959 &19.161 &19.150 &19.150 &19.150         \\
      &5$d_{5/2}$&-36.806&-36.092&-36.104&-36.104&-36.104        \\
\end{tabular}
\end{ruledtabular}
\end{table*}

 In the previous section we analyzed and compared our results with the 
earlier ones in some detail. Majority of our results are in very good agreement 
with the experimental data, some are perhaps the best match. The earlier works
chosen for comparison are based on diverse types of orbitals: numerical, 
finite discrete spectrum, B-spline and GTO. These are an accurate 
representation of the tried and tested types of single orbital in atomic 
calculations. Similarly, there is a variation in the many-body methods: MBPT, 
MCDF-EOL and coupled-cluster. This is a large data set for comparison. More
importantly, among the ions considered there are large changes in the role of 
electron correlations. This choice is essential to deliberate on the 
consequence of higher order terms and avoid erroneous inference from an 
incomplete sample. This sets the stage for a systematic appraisal of the 
higher order terms in the properties calculations. 

 As discussed in Section.\ref{all-order}, we implement the iterative method
to calculate the hyperfine constant to all orders for the {\em loe} one. To 
frame the iterative equation in terms of components, define $\tau$ as 
c numbers in the second quantized representation of ${\cal H}$. That is 
\begin{equation}
  {\cal H} = \sum_{ij} \tau _i^ja^\dagger_ia_j + \sum_{ijkl} \tau _{ij}^{kl}
             a^\dagger_ka^\dagger_l a_ja_i + \cdots.
\end{equation}
The Eq.(\ref{iter_eqn}) then assumes the form
\begin{eqnarray}
 \tau _a^p &= & h_{pa} + h_{pq}t_a^q + h_{ba}t_b^p + h_{bq}\tilde t_{ab}^{pq} + 
                h_{bq}t_a^qt_b^q +  \nonumber \\
           &  & \tau _b^q\tilde t_{bc}^{*qr} \tilde t_{ca}^{pr} + 
                \tau_c^p t_{bc}^{*qr}\tilde t_{ab}^{qr} + 
                \tau_a^r t_{bc}^{*qr}\tilde t_{bc}^{qp},
\end{eqnarray}
where $h_{ij}$ is the matrix element $\langle i|h_{\rm hfs}|j\rangle$ and 
$\tilde t_{ij}^{kl} = t_{ij}^{kl} - t_{ji}^{kl}$ is the antysymmetrised 
cluster amplitude. 
This is the equation we solve iteratively till convergence. After each 
iteration, we evaluate the contribution from the effective operator to the
hyperfine constant $S_2^\dagger {\cal H}_1$. The results of our calculations 
are given in Table.\ref{tab:all_order}.  For most of the cases, the 
results converges to KHz accuracy after two iterations. 

 In terms of absolute changes, the largest is observed in 
$6s\; ^2S_{1/2}$ of Ba$^+$. For this state the zeroth iteration, arising 
from ${\cal H}_1^0$, as given in Table.\ref{tab:all_order} is $469.636$. 
It converges to $467.450$ at the second iteration and change is $-2.186$. 
Which is 0.5\% of the zeroth iteration and 0.05\% of the total value. Whereas
in terms of fractional change, the largest is $5d\; ^2D_{5/2}$. Upon 
convergence the change is -0.702, which is 1.9\% of zeroth iteration. 
However, this is 5.5\% of the total value. This arises from the large
cancellation between the DF and $S^\dagger\tilde H_{\rm hfs}$. Here to obtain
correct result the iterated calculation should be applied to the other
terms as well. Not very surprisingly, the changes in Mg$^+$, Ca$^+$ and Sr$^+$ 
which have lower $Z$ are negligibly small. 

 Considering that iteration is implemented for the {\em loe} which 
contributes maximally. Contributions from the other {\em loe} is expected to
much smaller.

%%%%%%%%%%%%%%%%%%%%%%%%%%%%%%%%%%%%%%%%%%%%%%%%%%%%%%%%%%%%%%%%%%%%%%%%%%%%%%
%%%%%                  Section: Conclusions                             %%%%%%
%%%%%%%%%%%%%%%%%%%%%%%%%%%%%%%%%%%%%%%%%%%%%%%%%%%%%%%%%%%%%%%%%%%%%%%%%%%%%%

\section{Conclusions} 

 We have calculated, as well as surveyed and compared the magnetic hyperfine 
structure constants of low lying states of Mg$^+$, Ca$^+$, Sr$^+$ and Ba$^+$ 
available in the literature. For the states 
$3s\;^2S_{1/2}$ ($^{25}{\rm Mg}^+$), $3d\;^2D_{5/2}$ ($^{43}{\rm Ca}^+$), 
$4d\;^2D_{5/2}$ ($^{87}{\rm Sr}^+$) and $6s\;^2S_{1/2}$ ($^{137}{\rm Ba}^+$), 
our results provides the best match with the experimental data. Further more, 
results for most of the other states are in very good agreement with the 
available experimental data. 

  The chosen systems have hyperfine constants with varying dependence on
electron correlations. It is a suitable data set to examine the impact of 
higher order terms in properties calculations with relativistic coupled-cluster
theory. This is of paramount importance for high precision properties 
calculations with relativistic coupled-cluster. Our study establish 
without any ambiguity, the higher order terms are not important
when the leading terms DF and $S^\dagger\tilde H_{\rm hfs}$ contribute 
coherently. However, when large cancellation occurs like in $^2D_{5/2}$ state
of alkaline Earth ions, a consistent calculation of the different terms to 
equal orders is a must. Except for such cases, based on the present study, we 
recommend 
\begin{eqnarray}
  \langle \Psi_v| H_{\rm hfs}|\Psi_v\rangle &=& \langle\Phi_v| H_{\rm hfs} + 
             2H_{\rm hfs}T_1 + T_1^\dagger H_{\rm hfs}\left ( T_1 + 2T_2 
             \right ) \nonumber \\
           && + T_2^\dagger H_{\rm hfs}T_2 + 2S^\dagger 
              \left ( H_{\rm hfs}e^T\right )_1  \nonumber \\
           && + S_1^\dagger H_{\rm hfs}\left ( S_1 + 2S_2 \right ) + 
              S_2^\dagger H_{\rm hfs}S_2 |\Phi_v\rangle , 
\end{eqnarray}
to calculate hyperfine and similar properties for single valence systems. 
It is sufficient to include terms up to quadratic in $T$ for properties 
calculations. Higher order terms, all together, have  less than 
0.1\% of the total value and can be neglected.

%%%%%%%%%%%%%%%%%%%%%%%%%%%%%%%%%%%%%%%%%%%%%%%%%%%%%%%%%%%%%%%%%%%%%%%%%
%%%%%%%           Acknowledgements                                 %%%%%%
%%%%%%%%%%%%%%%%%%%%%%%%%%%%%%%%%%%%%%%%%%%%%%%%%%%%%%%%%%%%%%%%%%%%%%%%%

\begin{acknowledgments}
We wish to thank S. Chattopadhyay, S. Gautam, K. V. P. Latha, B. Sahoo and
S. A. Silotri  for useful discussions. DA gratefully acknowledges discussions 
with D. Mukherjee and B. P. Das. The results presented in the paper
are based on computations using the HPC cluster at Physical Research 
Laboratory, Ahmedabad.
\end{acknowledgments}

%%%%%%%%%%%%%%%%%%%%%%%%%%%%%%%%%%%%%%%%%%%%%%%%%%%%%%%%%%%%%%%%%%%%%%%%%%%%%%%
%%%%%%%%%%%                References                    %%%%%%%%%%%%%%%%%%%%%%
%%%%%%%%%%%%%%%%%%%%%%%%%%%%%%%%%%%%%%%%%%%%%%%%%%%%%%%%%%%%%%%%%%%%%%%%%%%%%%%

\end{document}